\documentstyle[11pt]{article}

\title{\vspace{4cm}\large\bf
Physical Vacuum and Cosmic Coincidence Problem }
\author{Arthur D.~Chernin\\
 Sternberg Astronomical Institute, Moscow University, Moscow, 119899,
Russia, \\
Astronomy Division, FIN-90014 University of Oulu, Finland,\\
and Tuorla Observatory, University of Turku, Piikki\"o, FIN-21500, Finland
}

\date{~}

\begin{document}

\maketitle

\begin{abstract}
\noindent A framework is suggested in which the energy integrals
of the Friedmann cosmology are identified as genuine
time-independent physical characteristics for both vacuum and
non-vacuum forms of cosmic energy. The integrals are found to be
numerically coincident within two orders of magnitude. It is
assumed that this coincidence reveals a symmetry that relates
vacuum to non-vacuum forms of cosmic energy at fundamental level.
The symmetry shows the cosmic coincidence problem and the
naturalness problem as two inter-related aspects of a more
general problem: Why are the energy integrals numerically
coincident and equal to $ \sim 10^{60} M_{Pl}^{-1}$? A simple
kinetics model of cosmological freeze out is used to examine how
-- at least, in principle -- the electroweak scale physics might
explain the nature of the symmetry between vacuum and non-vacuum
cosmic energies and determine the value of the energy integrals
in terms of the fundamental energy scales.

Keywords: Cosmology; Dark matter

PACS numbers: 04.70.Dy; 04.25.Dm; 04.60-m; 95.35.+d; 98.80.Cq

\end{abstract}
%
%
\section{Introduction}

The four major forms of energy in current cosmological models are
the acceleration (A) energy which is vacuum and/or a rolling
scalar quintessential field, cold dark (D) matter, baryons (B)
and ultra-relativistic (R) energy which includes cosmic microwave
background (CMB) photons and other possible light or massless
particles (neutrinos, gravitons, etc.) of cosmological origin.
The first of these forms of cosmic energy has recently been
discovered in the observations of high-redshift supernovae (Ries
et al. 1998, Perlmutter et al. 1999), and this has  been also
confirmed by evidence from the cosmic age, the large-scale
structure, the CMB anisotropy in combination with cluster
dynamics and  galaxy velocity field, etc. (see Cohn 1998, Carol
2000, Bahcall et al. 2000 and references therein).

The best  fit concordant with the bulk of all observational
evidence today is provided by the following figures for the
densities of these forms of energy (see the references above):
\begin{equation}
\Omega_A = 0.7 \pm 0.1, \;\; \Omega_D = 0.3 \pm 0.1, \;\;
\Omega_B = 0.02 \pm 0.01, \;\; \Omega_R = 0.6 \alpha \times
10^{-4},
\end{equation}
\noindent where $1 <\alpha < 30-100$ is a dimensionless constant
factor that accounts for non-CMB contributions to the relativistic energy.
With the Hubble constant $h_{100} = 0.65 \pm 0.15$, the figures
lead to either an open cosmological model or a spatially flat model.

The density of A-energy,  $\rho_A \sim 10^{-123} M_{Pl}^4$ (in the units
in which $c = k = \hbar =1; M_{Pl} = 1.2 \times 10^{19}$ GeV is the
Planck mass), conflicts drastically with simple field
theoretic expectations, if this density is treated as the density of
physical vacuum. Symmetry arguments may explain why the vacuum density
should be either zero or Planckian, but there is no explanation from the
`first principles' for a non-zero, tiny and positive vacuum density.
This is the well-known `naturalness problem in theoretical physics'
(Weinberg 1989).

On the other hand, the density of acceleration matter is near-coincident
with the density of dark matter and also with the densities of the
two other forms of non-vacuum energy,
as is seen from Eq.(1); this is another
problem, known as the `cosmic coincidence problem', that has been
recognized as a severe challenge to current cosmological concepts
(Wang et al. 2000).

One possible approach to both problems would be to assume that
the acceleration of the expansion is produced by non-vacuum energy  which
has negative pressure and negative effective gravitating
density (Peebles and Ratra 1988, Frieman and Waga 1998, Caldwwell and
Steinhardt 1998, Caldwell et al. 1998, Zlatev et al. 1999).
This energy form is called quintessence;
it can naturally be realized in some scalar field models in which the field
depends on time, and so the density of quintessence is diluted with
the  expansion. If so,  the densities involved  in the observed cosmic
coincidence are all diluted, which makes them seemingly more similar
to each other. Quintessence density might temporarily or even always be
comparable
to the density of dark matter. Although such an idea may make the
closeness of all cosmic densities natural, it does not explain the
coincidence that the quintessence field becomes settled with a finite energy
density comparable to the matter energy density just now (see, for instance,
Arkani-Hamed et al. 2000).

In the present paper, a framework is suggested that is
alternative to the idea of quintessence.
The cosmic acceleration is understood below as a dynamical effect of
physical vacuum with constant density $\rho_V = \rho_A$ in any reference
frame. In this framework, the cosmic coincidence problem and the naturalness
problem are treated as two basic aspects of a more fundamental problem
in theoretical physics and cosmology. This new problem can be formulated
 in terms of energy integrals of the Friedmann cosmology.

In Secs.2,3 it is demonstrated that the energy integrals are
constant genuine physical characteristics of both vacuum and
non-vacuum energy forms;
 on this basis, the formulation of the problem is given in Sec.4;
in Sec.5, it is argued that the electroweak scale physics can be
the major mediator in the interplay between vacuum and non-vacuum
energy forms, and a kinetics model that describes this process at
TeV temperatures in early Universe is studied in Sec.6; a brief
summary is given in Sec.7.

\section{Energy integrals}

The constant energy integrals  enter the general Friedmann solution for the
four cosmic energies mentioned above:
\begin{equation}
\int{da (A_V^{-2} a^2 + 2 A_D a^{-1} + 2A_B a^{-1} +
A_R^2 a^{-2} - K)^{-1/2}} = t,
\end{equation}
\noindent were $a(t)$ is the curvature radius and/or a scale
factor of the model, $K= 1,0,-1$, accordingly to the sign of the
spatial curvature. Constants $A$ come from the Friedmann
`thermodynamic' equation which is equivalent to energy and
entropy conservation relation in a co-moving volume during the
process of adiabatic expansion:
\begin{equation}
A = [ (\frac{1 + 3w}{2})^{2} \; \kappa \rho a^{3(1+w)}]^{\frac{1}{1 + 3w}}.
\end{equation}
Here $w = p/\rho$ is the pressure-to-density ratio for a given energy form;
$w = -1, 0, 0, 1/3$ for vacuum, dark matter, baryons and radiation,
respectively; $\kappa = 8\pi G/3 = (8\pi/3) M_{Pl}^{-2}$.
The interpretation of the constants $A_D, A_B, A_R$
 is quite obvious in Eq.(3) for non-vacuum energies: they
express the conservation of the number of particles in the
co-moving volume. It is most interesting
 that the constant for vacuum $A_V$ is also given by the same general
 relation of Eq.(3), while
 the interpretation in terms of particles does not work in this later case.

In an explicit form, the solution of Eq.(3) may be written
separately for the earlier epoch ($z > z_V \simeq 1$), when the
non-vacuum energies dominate and for the later vacuum domination
epoch ($z < z_V$). For a spatially flat model at $z > z_V$ one
has (Chernin 1965)
\begin{equation}
a(\eta) = A_M \eta ^2 + A_R \eta, \;\;\; t(\eta) = \frac{1}{3} A_M
\eta ^3 + \frac{1 }{2}A_R \eta ^2.
\end{equation}
\noindent
Here $\eta$ is conformal time, and $A_M, A_R$ are given
by Eq.(3) with $\rho = \rho_M = \rho_D + \rho_B, w = 0$ and $\rho =
\rho_R, w = 1/3$, respectively.

For the vacuum domination epoch, one has a well-known solution that describes
accelerating expansion controlled by vacuum only:
\begin{equation}
a(t) = A_V f(t), \;\;\; f(t) = \sinh (t/A_V), \;\; \exp (t/A_V),
\;\; \cosh (t/A_V),
\end{equation}
\noindent
for open, spatially flat and close models, respectively. The present-day
($t = t_0 \simeq 15 \pm 2$ Gyr)  value of $a(t)$ is
estimated with this solution and the observed Hubble constant;
one has approximately: $a(t_0) \sim A_V$ for all the three models.

Eqs.(2-5) suggest a framework in which the time-independent integrals
for all the four energy forms are considered as basic cosmological
quantities.

Indeed, the energy integrals given by Eq.(3) are genuine constant
characteristics for the respective forms of energy during all the
time when each of the energies exists (and not only at the epoch
when a given energy dominates -- see the solutions of Eqs.2,4,5
above). Each of the energies is represented by its corresponding
integral independently from other components. Being constants of
integration, the integrals are completely arbitrary, in the sense
that the Friedmann equations  provide no limitations on them,
except for trivial ones. From the view point of physics, the
integrals are determined by `initial conditions' at the epoch of
the origin of the forms of energy in the early Universe; at that
time, each of the energies  acquires its own integral $A$  as a
genuine quantitative characteristic, which is then kept constant
in time.

\section{Evaluation of energy integrals}

The numerical estimation of the integrals can be made with the figures
of Eq.(1) and with the use of Eqs.(3,5):
\begin{equation}
 A_V = (\kappa \rho _A)^{-1/2} \sim 10^{42} GeV^{-1} \sim 10^{61} M_{Pl}^{-1},
\end{equation}
\begin{equation}
A_D = \frac{1}{4} \kappa \rho_D a^3 \sim 10^{41} GeV^{-1}
\sim10^{60} M_{Pl}^{-1},
\end{equation}
\begin{equation}
 A_B = \frac{1}{4} \kappa \rho_B a^3
\sim 10^{40} GeV^{-1} \sim 10^{59} M_{Pl}^{-1},
\end{equation}
\begin{equation}
A_R = (\kappa \rho _R)^{1/2} a^2 \simeq 10^{40} GeV^{-1}
 \sim 10^{59} M_{Pl}^{-1}.
\end{equation}

The integrals (that have the dimension of the length) are
evaluated for the open model with $a(t_0)$ given by Eq.(5). It is
easy to see that the evaluation for the flat (with the scale
factor normalized as it is indicated above in Eq.(5)) and close
models gives rise to similar results, on the order of magnitude.
In the estimation of $A_R$, a conservative value $\alpha = 1$ is
adopted which takes into account the CMB photons only.

As one sees, the integrals have proven
to be close to each other within two orders of magnitude, and this result
may be summarized in a compact formula:
\begin{equation}
A  \sim 10^{60 \pm 1 } M_{Pl}^{-1}.
\end{equation}

Note that the analysis may easily be extended to include quintessence
alongside with or instead of vacuum. For instance, according to Eq.(3),
quintessence with $w = - 2/3$ is characterized by $A_{Q} \sim
10^{61} M_{Pl}^{-1}$, if it is estimated with $\rho_q = \rho_A$ from Eq.(1).
A special case is the integral for non-accelerating energy
 with $w = - 1/3$ which is not contained in Eq.(3);  using directly the
 corresponding solution of the Friedmann equations, one finds:
 $A_{-1/3} = A_V$, if $\rho_{-1/3} = \rho_A$.

\section{Symmetry of cosmic energy forms}

The integrals  of Eqs.(3,10) exist in
the Universe since the epoch at which the four major forms of
energy came to existence themselves, e.g. at least after $t \sim 1$ sec,
and will exist until the
decay of the protons at $t \ge 10^{31-32}$ yrs or the decay of
the particles of dark matter. In the beginning of this time
interval, the vacuum density is about forty orders of magnitude
less than the density of R-energy that dominates at that epoch;
but the numbers $A_A$ and $A_R$ are as close at $t \sim 1$ sec
as they are in Eqs.(6,9). In the future, the scale factor will change
in  orders of of orders (!) of magnitude during the life-time
of proton, and so the densities of D-, B-, and R-energies, as
well as their ratios to the vacuum density, will
change enormously. But the four constant numbers of Eq.(6-10) will remain the
same keeping their near-coincidence for all this future time.

The empirical analysis of Secs. 2,3 has led to a novel version of
the cosmic coincidences that appear now as a tetramerous
coincidence of time-independent constant numbers -- in contrast
to the epoch-dependent (and therefore temporal, and accidental,
in this sense) coincidence of densities. It is clear that the
densities are coincident at the present epoch because this is the
epoch of the transition from the decelerated matter dominated
expansion to the accelerated vacuum domination expansion. Indeed,
since the relation of Eq.10 exists from the early state of the
Universe, the densities must be coincident just now, because
$a(t) \sim A_V$ presently. Another question is why do we happen
to live in a transition epoch; this is among the matters that may
be discussed with antropic principle (see,for instance, Weinberg
1987).

The coincidence of the four integrals reveals a new type of basic regularities
which are conserved in the evolving Universe.
Since none of the energy forms looks preferable, in terms of the energy
integrals of Eq.10, the regularity has a character of a symmetry that
relates vacuum
to  non-vacuum forms of cosmic energy. The symmetry is not perfect, and
its accuracy
is within a few percent, on logarithmic scale, according to Eqs.(6-10).

The symmetry of two forms of energy, namely B- and
R-energies, in terms of energy integrals, was first recognized
soon after the discovery
of the CMB (Chernin 1968). The significance of this B-R symmetry is obvious
from the fact that the relation
$A_B \sim A_R  \sim 10^{59} M_{Pl}^{-1}$ enables alone to quantify
such important things as the baryon-antibaryon asymmetry
of the Universe, the entropy per baryon,
the light-element production in the nucleosynthesis, the epoch
of hydrogen recombination, the present-day temperature of CMB,
etc.

In the framework of the symmetry extended to the all vacuum and
non-vacuum cosmic energies, the genuine constant characteristic
of physical vacuum $A_V$ finds its natural non-vacuum
counterparts which are also constant in time. Since all the forms
of cosmic energy are treated in a unified manner and on the
common ground of the energy integrals, the naturalness problem
looses its uniqueness and finds its place in this broader
cosmological framework. This way the naturalness problem together
with the cosmic coincidence problem are transformed into a more
general problem in fundamental physics and cosmology: Why is $A_V
\sim A_D \sim A_B \sim A_R \sim 10^{60} M_{Pl}^{-1}$?

\section{Vacuum and electroweak scale physics}

In any fundamental unified theory, the value of the vacuum density would
have to be calculable. Alongside with this,
the symmetry of energy forms described by Eq.10 would have to be explained.
This seems to be a rather remote goal. What could seemingly be done now
is to try, first,
to identify a clue physical factor that determines both the symmetry and
the value of the vacuum density.

One possible way to do this is to assume that the vacuum density
is completely due to zero oscillations of the quantum fields, as
it was suggested not once since the 1930-s (see, for instance,
Dolgov et al. 1988). If so, the vacuum density is given by an
integral over all frequencies of the zero oscillations, -- but
this gives infinite value for the density. One may cut off the
range of the frequencies, introducing a maximal frequency
$\omega_V$; then the integral takes a form: $\rho_V \sim \omega_V
^4 $ (as it can be seen, for instance, from dimension
considerations). With the observed vacuum density, one finds from
this that $\omega_V \sim 10^{-31} M_{Pl}$. Then one may put this
maximal frequency into the cosmological context and compare it
with the rate of cosmological expansion in the early Universe
$1/t \sim (G \rho)^{1/2}$;  here $\rho \sim \rho_R \sim T^4$, and
$T$ is the temperature of radiation which dominates at that
times. From this one finds that $\omega_V \sim 1/t$ at the epoch,
when $T \sim 10^{-16} M_{Pl} \simeq 1$ TeV. The last value is
close to the electroweak energy scale $M_{EW}$, and so $\omega_V
\sim M_{EW}^2/M_{Pl}$.

With this maximal frequency, the vacuum density is
\begin{equation}
\rho_V \sim \omega_V^4 \sim (M_{EW}/M_{Pl})^8 M_{Pl}^4.
\end{equation}

These considerations suggest that the electroweak scale physics
might be behind the observed density of cosmic vacuum.  This
physics may also be the major mediator between vacuum and
non-vacuum forms of cosmic energies in the processes that might
develop in the early Universe at the epoch of TeV temperatures.

Note that the relation of Eq.(11) was obtained also in a
field-theoretic model (Arkani-Hamed 2000, -- unfortunately, under
some  arbitrary additional assumptions).

\section{A model}

Having in mind the central part that may be played by the electroweak scale
physics at the epoch of TeV temperatures, one may try to examine, if this
physics could determine also the symmetry between vacuum and non-vacuum
energies. For this purposes, a kinetics model can be used, in which
non-relativistic dark matter is considered as thermal relic of early
cosmic evolution. This possibility has been widely discussed (see, for
instance, the books by Zeldovich and Novikov 1983, Dolgov et al. 1988,
Kolb and Turner 1990, and a recent work by Arkani-Hamed et al. 2000).
The model below is incomplete: baryonic energy is not included in it,
 and so baryogenesis at TeV temperatures must be
studied separately, but, perhaps, not independently of the model.
As for vacuum, dark matter and radiation,
they will be represented in the model by the corresponding energy
integrals.

For stable (or long-living) particles
of the mass $m$, the abundance freezes out when the temperature falls below
the mass $m$ and the expansion rate $1/t$ wins over the  annihilation rate,
$\sigma n$, where the annihilation cross-section $\sigma \sim m^{-2}$. So
that at that moment the particle density is
\begin{equation}
n \sim 1/(\sigma t) \sim m^2 (G \rho_R).
\end{equation}
\noindent
Using Eq.(3) for $ A_D$ and $A_R$ and putting  $\rho_D \sim m n$,
one finds:
\begin{equation}
A_D \sim a(t) m^3 M_{Pl}^{-2} A_R.
\end{equation}
\noindent
One also has at that moment $\rho_R \sim m^4$, and because of this
\begin{equation}
 A_R \sim  a(t)^2 m^2 M_{Pl}^{-1},
\end{equation}
where $a(t) \sim A_V (1 + z)^{-1}$, and $z$ is the redshift
at the freeze-out epoch;
in this way, the vacuum integral comes to the model.

To specify the underlying fundamental physics, one
can refer to a special significance
of the electroweak scale physics, as it was mentioned above, and assume that
only two fundamental energy scales are involved in the process, namely
$M_{EW}$ and $M_{Pl}$. If so, it is natural to
identify the mass $m$ with the electroweak energy scale
$M_{EW}$. Then, one may use the relation for the
vacuum density in terms of the two fundamental energy scales,
$M_{EW}$ and $M_{Pl}$, as given by Eq.(11).  With this density, the vacuum
integral is
\begin{equation}
A_V \sim  (M_{Pl}/M_{EW})^4 M_{Pl}^{-1}.
\end{equation}

Arguing along this line, one may expect that
the redshift $z$ at the freeze-out epoch may be
a simple combination of the same two mass scales:
\begin{equation}
z \sim M_{Pl}/M_{EW}.
\end{equation}

Now the kinetics model is described by a system of four algebraic Eqs.(13-16)
(with $m = M_{EW}$) for the four numbers $A_M, A_R, A_V, z$. The solution
of the system is:
\begin{equation}
A_M \sim A_R \sim A_V \sim  (M_{Pl}/M_{EW})^4 M_{Pl}^{-1}.
\end{equation}

Thus the coincidence of the three energy integrals appears as a direct result
of the freeze-out process mediated by the electroweak scale physics.
This physics determines also the value of the integrals.

Following Arkani-Hamed (2000) and another recent work by Kawasaki
et al. (2000),  one may introduce  the gravitational scale $M_G$,
or the reduced Planck scale $m_{Pl}$, instead of the standard
Planck scale: $M_G \simeq m_{Pl} \simeq g M_{Pl}$, where $g
\simeq 0.1-0.3$. The dimensionless factor $g$ accounts for the
fact that the gravity constant $G$  enters the exact cosmology
relations in combinations like $ 8 \pi G/3,  6\pi G$, or  $32 \pi
G/3$. Similarly, a few dimensionless factors, like the effective
number of degrees of freedom, etc., may also be included in the
model  -- see again the books mentioned above. One gets finally:
\begin{equation}
 A \sim g^4 (M_{Pl}/M_{EW})^4 M_{Pl}^{-1} \sim 10^{61 \pm 1} M_{Pl}^{-1},
\;\;\; w = [-1, 0, 1/3].
\end{equation}
\noindent
A quantitative agreement with the empirical result of Eq.(10) looks
satisfactory here.
With $A_V$ of Eq.(15), one finds that the vacuum density is
$\rho_A \sim g^8 (M_{Pl}/M_{EW})^8 M_{Pl}^4 \sim 10^{-122 \pm 2} M_{Pl}^4$.

The numerical value of the redshift in Eq. (16)
$z \sim g M_{Pl}/M_{EW} \sim 10^{15}$, and so the
temperature at the freeze-out is $T \sim 1$ TeV $\sim M_{EW}$, which
reflects once again the central role of the electroweak energy scale
in the model above.

Thus, the model shows how -- at least, in principle -- the
problem of the symmetry of cosmic energy forms can be discussed in terms of
basic physics. The model gives an example of a possible
(while partial and incomplete yet) solution to this problem.

\section{Conclusions}

The results of the discussion above may be summarized as follows:

1. A framework is suggested in which the energy integrals of the Friedmann
cosmology equations (see Eqs.2,3) are
identified as genuine time-independent physical characteristics for both
vacuum and non-vacuum forms of cosmic energy.

2. The empirical analysis of the observational data on the four major
energy forms leads to the conclusion that the corresponding energy
integrals are numerically coincident within two orders of magnitude
(Eqs.6-10).

3. It is argued that the coincidence of the energy integrals found at the
empirical level reveals a symmetry that relates vacuum to
non-vacuum forms of cosmic energy at fundamental level.

4. The symmetry shows the cosmic coincidence problem and the
naturalness problem as two inter-related aspects of a more
general problem: Why are the energy integrals numerically
coincident and equal to $ \sim 10^{60} M_{Pl}^{-1}$?

5. A special significance of the electroweak scale physics is
demonstrated for this general problem.

6. Under the assumption that dark matter is a thermal relic of the early
Universe, a simple kinetics model of cosmological freeze out is used to
demonstrate how -- at least, in principle -- the electroweak scale physics
might explain the nature of the symmetry between vacuum and non-vacuum
cosmic energies and determine the value of the energy integrals in terms
of the fundamental energy scales.

The work was partly supported by the grant of the Academy of Finland
`Galaxy streams and dark matter structures'.
\vspace{0.5cm}

\section*{References}

Arkani-Hamed, N., Hall, L.J., Kolda, Ch., H. Murayama, H., 2000.
Phys. Rev. Lett. 85, 4434.

Bahcall, N., Ostriker, J., Perlmutter, S., and Steinhard, P.,
1999. Science 284, 1481.

Caldwell, R.R., and Steinhardt, P.J., 1998. Phys.Rev. D 57, 6057.

Caldwell, R.R., Dav\'e, R., and Steinhardt, P.J., 1998. Phys.Rev. Lett.
80, 1582.

Carol, S., 2000. Astro-ph/0004075.

Chernin, A.D., 1965. Sov. Astron. 42, 1124.

Chernin, A.D., 1968. Nature (London) 220, 250.

Cohn, J., 1998. Astro-ph/9807128.

Dolgov, A.D., Zeldovich, Ya.B., Sazhin, M.V., 1988.
{\em Cosmology of the Early Universe.}
(In Russian; Moscow Univ. Press, Moscow.

Frieman, J.A., Waga, I., 1998. Phys. Rev. D 57, 4642.

Kawasaki, M., Yamaguchi, M., Yanagida, T., 2000. Phys. Rev. Lett. 85, 3572.

Kolb, E.W., Turner, M.S., 1990.
{\em The Early Universe.} (Addison-Wesley, Reading).

Peebles, P.J.E., Ratra, B., 1988. Astrophys. J. Lett. 325, L17.

Perlmutter, S. {\it et al.}, 1999.  Astrophys. J. 517, 565.

Riess, A.G. {\it et al.},  1998. Astron. J. 116, 1009.

Wang, L., Caldwell, R.R., Ostriker, J.P., and Steinhardt, P.J., 2000.
Astrophis. J. 530, 17.

Weinberg, S., 1987. Phys. Rev. Lett. 61, 1.

Weinberg, S., 1989. Rev. Mod. Phys. 61, 1.

Zeldovich, Ya.B., Novikov, I.D., 1983. {\em The Structure and Evolution of the
Universe.} (The Univ. Chicago Press, Chicago and London).

Zlatev, I., Wang, L., Steinhard, P.J., 1999. Phys. Rev. Lett. 82, 896.

\end{document}